\documentstyle[aps,prl,preprint]{revtex}

\begin{document}

\draft

\title{
Characterization of the transition from defect- to
phase-turbulence}

\author{
David~A.\ Egolf\cite{CNCS-address}\cite{dae-email} and
Henry~S. Greenside\cite{CNCS-address}\cite{CS-Dept}
}

\address{
Department of Physics\\
Duke University, Durham, NC 27708-0305
}

\date{July 19, 1994}

\maketitle

\begin{abstract}

For the complex Ginzburg-Landau equation on a large
periodic interval, we show that the transition from
defect- to phase-turbulence is more accurately
described as a smooth crossover rather than as a sharp
continuous transition.  We obtain this conclusion by
using a powerful parallel computer to calculate various
order parameters, especially the density of space-time
defects, the Lyapunov dimension density, and the
correlation lengths of the field phase and amplitude.
Remarkably, the correlation length of the field
amplitude is, within a constant factor, equal to the
length scale defined by the dimension density. This
suggests that a correlation measurement may suffice to
estimate the fractal dimension of some large
homogeneous chaotic systems.

\end{abstract}

\pacs{
47.27.Cn,  
47.27.-i,  
47.27.Cn,  
05.45.+b,  
05.70.Ln,  
82.40.Bj   
}

\narrowtext

Recent advances in laboratory technique
\cite{space-time-expts} and in computer simulation
\cite{Shraiman92,Egolf94nature,ChateRefs,simulation-refs}
have opened up the study of boundary-independent
spatiotemporal chaos in large homogeneous sustained
nonequilibrium systems \cite{CrossHohenberg94}.  Many
fundamental questions remain unanswered about such
chaotic systems, e.g., what different states can occur,
how transport depends on different states, and what
kinds of bifurcations separate one state from another.
An especially interesting question is whether ideas
from statistical mechanics might be applicable to
chaotic nonequilibrium systems in the thermodynamic
limit of infinite system size
\cite{ThermoRefs,Miller93,Cross93}.

A significant step towards understanding some of these
questions was recently reported by Shraiman et al
\cite{Shraiman92}. These researchers studied different
spatiotemporal chaotic states of the one-dimensional
complex Ginzburg-Landau equation
\begin{equation}
  \partial_t u(x,t) =
      u
  +  (1 + i c_1) \partial^2_x u
  -  (1 - i c_3) |u|^2 u
  , \qquad \mbox{$x \in [0,L]$}
  , \label{1d-cgl}
\end{equation}
on a large periodic interval of length~$L=1024$, which
they assumed to be large enough to approximate the
thermodynamic limit of infinite system size.  Here the
variables~$t$ and~$x$ denote time and position
respectively, the complex-valued field~$u(x,t) = \rho
e^{i \phi}$ has magnitude~$\rho(x,t)$ and
phase~$\phi(x,t)$, and the parameters~$c_1 > 0$
and~$c_3 > 0$ are real-valued.  Eq.~(\protect\ref{1d-cgl}) is an
important model of spatiotemporal chaos because it is
simple, experimentally relevant \cite{cgl-expts} and
universal \cite{Cross93}: spatially extended systems
that undergo a supercritical Hopf bifurcation from a
static to oscillatory homogeneous state reduce to
Eq.~(\protect\ref{1d-cgl}) sufficiently close to the onset of the
bifurcation.  Interesting dynamics are predicted and
are observed beyond the Newell line $c_1 c_3 = 1$ since
all plane wave solutions of Eq.~(\protect\ref{1d-cgl}) are linearly
unstable to the Benjamin-Feir instability for $c_1 c_3
> 1$ \cite{Cross93}.

Shraiman et al summarized their simulations in the form
of a phase diagram in the $c_1$-$c_3$ parameter plane
(Fig.~3 of Ref.~\cite{Shraiman92}).  Based mainly on
calculations of the density of space-time defects~$n_D$
\cite{defect-phase-chaos-defns}, this diagram showed
continuous and discontinuous transition lines
(analogous to second- and first-order phase
transitions) separating defect-turbulent from
phase-turbulent states \cite{defect-phase-chaos-defns}.
Of special interest to us is the continuous
chaos-to-chaos transition line~$L_1$ in their Fig.~3,
which occurs for~$c_1 \ge 1.8$.  It is somewhat
mysterious why the density~$n_D$ decreases to zero at
an $L_1$~line that is distinct from the Newell line: in
the limit of infinite system size and of infinite time,
what prevents defects from forming anywhere to the
right of the Newell line ($c_3 > 1/c_1$)? The mystery
of the~$L_1$ line can be partly appreciated by trying
to reason by analogy to equilibrium statistical
physics. Assuming that the chaotic fluctuations of
Eq.~(\protect\ref{1d-cgl}) act as a finite-temperature ergodic
noise bath and observing that the derivatives in
Eq.~(\protect\ref{1d-cgl}) represent short-ranged interactions
between different parts of the field~$u$, we would not
expect distinct phases at finite temperature in
one-space dimension \cite{Landau80}.

Because so little is known about possible critical
phenomena of large homogeneous nonequilibrium systems
and because Eq.~(\protect\ref{1d-cgl}) is such an important model,
we have tried to characterize more carefully the
dynamics near the $L_1$~line for the fixed parameter
value~$c_1=3.5$.  By calculating various order
parameters over length scales as large as $L \le 10^6$
and over time scales as large as $T \le 10^7$, we are
able to show below that the change from defect- to
phase-turbulence near the $L_1$~line is more accurately
described as a smooth crossover rather than as a sharp
continuous transition with power-law scaling of order
parameters \cite{Shraiman92}. It is then a possibility
that phase turbulence (i.e., a chaotic state with~$n_D
= 0$ in the thermodynamic limit) does not exist
although we can not settle this with our present
computer resources. Our calculations also confirm
several points predicted by Shraiman et al and by other
researchers \cite{Grinstein94}, e.g., that the spatial
correlation function of the phase should decay
exponentially inside the phase-turbulent regime. We
have also studied whether the dimension
density~$\delta$ (Lyapunov fractal dimension per unit
volume) is a useful order parameter for characterizing
changes in spatiotemporal chaotic states
\cite{Cross93,Egolf94nature,OHern94}.  The dimension
density defines a dimension correlation length
$\xi_\delta = \delta^{-1/d}$ \cite{Cross93} which is
the characteristic size of dynamically independent
subsystems of spatial dimensionality~$d$
\cite{Egolf94nature}. A comparison of~$\xi_\delta$ with
other characteristic length scales as a function of the
parameter~$c_3$ gives the remarkable result
that~$\xi_\delta$ is, up to a constant factor, equal to
the spatial correlation length of the field magnitude,
$\xi_\rho$, from the Newell~line to beyond the
$L_1$~line.

An important resource for the advances reported below
was a CM-5 parallel computer \cite{Hillis93}, which
facilitated the study of much larger space-time and
parameter regions than were previously conveniently
accessible.  Our numerical methods for integrating
Eq.~(\protect\ref{1d-cgl}) and for calculating related order
parameters are similar to previous methods
\cite{Shraiman92} except for modifications of
algorithms and codes to take advantage of the CM-5's
scalable parallel architecture \cite{Egolf94nature}.
For most of our simulations, we used a time step
$\Delta{t}=0.05$ and a spatial resolution of two
Fourier modes per unit length; this space-time
resolution was dictated by the need to detect isolated
space-time defects when estimating the density~$n_D$
\cite{Shraiman92}. Our initial condition for most runs
was $u(x,t=0)=0.4 + \eta$ where~$\eta$ was a uniformly
distributed $\delta$-function correlated random
variable in the interval $[-0.02,0.02]$. We typically
averaged over~64 or more random initial conditions
spanning an integration time of $2 \times 10^5$ so that
the total effective integration time was perhaps as
long as $64 \times (2 \times 10^5) = 1.3 \times 10^7$
time units.

Before discussing our results, we note that
for~$c_1=3.5$, for an integration time of $T=10^5$, and
for a periodic interval of length~$L=1024$, Shraiman et
al argued the existence of the $L_1$~line using two key
observations \cite{Shraiman92}: (1), that the
density~$n_D$ vanished as a power law $n_D \propto (c_3
- \bar{c_3})^\alpha$ with exponent~$\alpha \approx 2$
and with $\bar{c_3} \approx 0.77 > c_3^{\text Newell} =
1/c_1 = 0.286$; and (2), that the correlation
time~$\tau$ of phase fluctuations \cite{tau-defn}
diverged as a power law also at~$\bar{c_3}$, as the
inverse of the defect density, $\tau \propto 1/n_D$.
For $c_3 < \bar{c_3}$, Shraiman et al observed a
less-disordered phase-turbulent regime with~$n_D$
empirically equal to zero and with
slower-than-exponential decay of temporal correlations
\cite{Shraiman92}. If defects do not occur in the
thermodynamic limit, a perturbation theory in the small
quantity $\epsilon = c_1 c_3 - 1$ yields a simpler
description of phase turbulence near the Newell line,
$\epsilon \to 0$. In that limit, Eq.~(\protect\ref{1d-cgl}) reduces
to the Kuramoto-Sivashinsky (KS) equation
\cite{KuramotoRefs,Cross93}
\begin{equation}
  \partial_t \phi =
   - \epsilon \, \partial_x^2{\phi}
   - {1 \over 2} c_1^2 ( 1 + c_3^2 ) \, \partial_x^4{\phi}
   - (c_1 + c_3) \, ( \partial_x{\phi} )^2
  , \qquad \epsilon = c_1 c_3 - 1
  , \label{ks-eq}
\end{equation}
and the amplitude~$\rho$ becomes an algebraic function
of a spatial derivative of the phase,
\begin{equation}
  \rho  \approx
    1  -  {c_1 \over 2} \partial_x^2\phi
  . \label{adiabatic-approximation}
\end{equation}
Some of our calculations below provide the first
quantitative comparisons of phase turbulence as
described by Eqs.~(\ref{ks-eq})
and~(\ref{adiabatic-approximation}) with phase
turbulence as empirically observed in Eq.~(\protect\ref{1d-cgl}).

For the parameter value~$c_1=3.5$, for a system
size~$L=4096$, and for an effective integration time
of~$T=10^7$ (after allowing transients of duration
$10^4$ to decay), we find in
Fig.~\protect\ref{fig:defect-density} that~$n_D$ is finite
substantially to the left of the $L_1$~line as
calculated in Ref.~\cite{Shraiman92}.  Far to the right
side of the $L_1$~line, our data in
Fig.~\protect\ref{fig:defect-density}(a) approximately reproduce
the previously reported \cite{Shraiman92} power-law
scaling with exponent~$\alpha \approx 2$. Closer to the
$L_1$~line, a least-squares fit of the three-parameter
expression $a (c_3 - \bar{c_3}')^\alpha$ to the nine
left-most points gives a much larger exponent $\alpha
\approx 6.8$, with an onset of phase turbulence ($n_D =
0$) at $\bar{c_3}'=0.74 < \bar{c_3}=0.77$.  Assuming
equal errors bars on all data points, we find the
chi-square value for the fit to be $\chi^2=4.6 \times
10^{-12}$.  The increase in the exponent with increased
space-time resolution suggests that a power-law scaling
is inappropriate. As shown in
Fig.~\protect\ref{fig:defect-density}(b), we find a better fit of
the same data with the functional form
\begin{equation}
  n_D  =
   a \exp\left(
     - b / \left(c_3 - \bar{c_3}'' \right)^\alpha
 \right) , \label{essential-singularity}
\end{equation}
which is the expected behavior for thermodynamic
Gaussian fluctuations of the phase
gradient~$\partial_x\phi$ if large values of the latter
are the reason for defect nucleation \cite{Shraiman92}.
If we set $\alpha=1$, a least-squares fit of
Eq.~(\protect\ref{essential-singularity}) to the nine left-most
data points yields the three parameter values $a=0.66$,
$b=0.98$, and $\bar{c_3}'' = 0.70 <
\bar{c_3}'=0.74$ with a $\chi^2 = 8.1 \times 10^{-13}$.
If we set $\bar{c_3}'' = c_3^{\text Newell} = 0.286$ to
test whether Eq.~(\protect\ref{essential-singularity}) is
consistent with the onset of phase turbulence at the
Newell line, a least-squares fit (again to the nine
left-most points) gives the parameter values $a=0.018$,
$b=0.017$, and $\alpha=8.8$ with a substantially poorer
$\chi^2=8.2 \times 10^{-11}$.  Our data spanning the
crossover region evidently lie too far to the right of
the Newell line to determine whether the defect density
goes to zero before or at this line.

To test independently the important implication of
Fig.~\protect\ref{fig:defect-density} that a crossover occurs (so
that phase turbulence may not be a distinct phase from
defect turbulence), we have calculated other order
parameters over the same parameter range. The
occurrence of a crossover is supported by
Fig.~\protect\ref{fig:correlation-times}, which summarizes
correlation times~$\tau$ of the phase~$\phi(x,t)$
\cite{tau-defn} as a function of~$c_3$.
Fig.~\protect\ref{fig:correlation-times}(a) shows that~$\tau$
does not diverge to infinity at the $L_1$~line (as
reported in Ref.~\cite{Shraiman92}) but instead is
large and finite a little bit to the left of the
$L_1$~line. We find that time correlation functions of
the phase decay exponentially down to at least~$c_3 =
0.77$ but do not have a simple functional form for
smaller~$c_3$ values. Recent arguments and calculations
have been made \cite{Grinstein94} that time correlation
functions should decrease as a stretched exponential
for the phase-turbulent regime of Eq.~(\protect\ref{1d-cgl}) in
which case a correlation time can not be meaningfully
defined.  However, the change from exponential to
stretched exponential behavior with decreasing~$c_3$
has not yet been carefully studied.  Shraiman et al
conjectured that the critical scaling of $\tau$ at the
$L_1$~line followed from the scaling of~$n_D$
with~$\tau \propto 1/n_D$.
Fig.~\protect\ref{fig:correlation-times}(b) shows that this is
approximately correct to the right of the $L_1$~line
but breaks down closer to this line since the
product~$\tau n_D$ is no longer constant: the defect
density goes to zero faster than the correlation time
increases.

In Fig.~\protect\ref{fig:correlation-lengths}(a), we have
calculated the phase spatial correlation length
$\xi_\phi$ \cite{tau-defn} as a function of~$c_3$.
Shraiman et al argued that~$\xi_\phi$ should be finite
in the phase turbulent regime of Eq.~(\protect\ref{1d-cgl}) and
estimated its value indirectly by calculating a phase
diffusion coefficient~$D=1/\xi_\phi$ from
phase-gradient correlations \cite{Shraiman92}.
Exponential decay of spatial correlations is also
expected for phase turbulence if the latter is
described at long-wavelengths by the
Kardar-Parisi-Zhang (KPZ) Langevin equation
\cite{Grinstein94}.  By going to quite large system
sizes ($L = 10^6$) and to long integration times, we
have verified directly that the phase spatial
correlation function \cite{tau-defn} decays
exponentially well to the left of the $L_1$-line as
shown in Fig.~\protect\ref{fig:correlation-lengths}(a). As the
parameter~$c_3$ decreases, the quantity~$\xi_\phi$
varies {\em smoothly} through a local maximum near the
$L_1$~line, and then increases steadily until we can no
longer estimate its value accurately with our computer
resources.  The smooth variation of~$\xi_\phi$ through
the $L_1$~region is consistent with a crossover rather
than with a sharp transition.  The apparent divergence
of~$\xi_\phi$ upon approaching the Newell line,
$\epsilon \to 0$, can be understood semiquantitatively
as shown in Fig.~\protect\ref{fig:correlation-lengths}(a) by a
scaling argument \cite{xi-phi-scaling} that predicts
$\xi_\phi \propto \epsilon^{-5/2}$. The agreement is
within about 10\%.

The phase correlation length~$\xi_\phi$ is the same as
that of the field~$u$ itself \cite{Egolf94nature}, but
there is a separate, generally shorter, correlation
length scale~$\xi_\rho$ associated with fluctuations of
the field amplitude~$\rho$ (also with the phase
gradient~$\partial_x\phi$).
Fig.~\protect\ref{fig:correlation-lengths}(b) compares the
reciprocals of the phase and amplitude correlation
lengths with the Lyapunov dimension density~$\delta$,
whose reciprocal defines the dimension correlation
length $\xi_\delta$ discussed above
\cite{Egolf94nature}.  Up to constant factor of~$1.4$,
we find that the amplitude correlation length equals
the dimension correlation length $\xi_\delta$ over a
substantial range of parameter~$c_3$. (An independent
and related result was also recently reported by other
researchers \cite{Bohr94comment}.)  {\em This
remarkable result suggests that the big fractal
dimension of some large homogeneous chaotic systems
might be accurately estimated by simple correlation
function calculations.} In work that we will discuss
elsewhere \cite{OHern94}, we have also calculated the
variation of~$\xi_\delta$ across a nonequilibrium Ising
transition in a two-dimensional coupled map lattice
\cite{Miller93}.  Although the agreement is not quite
so striking, $\xi_\delta$ still matches closely the
correlation length associated with fluctuations of the
magnitude of the site variables.

In Fig.~\protect\ref{fig:delta-and-rms}, we make two final
comparisons of how phase-turbulence, as described by
the adiabatic approximation
Eq.~(\protect\ref{adiabatic-approximation}) and by solutions of the
KS-equation Eq.~(\protect\ref{ks-eq}), agrees with numerical
solutions of Eq.~(\protect\ref{1d-cgl}). The dimension
density~$\delta$ of the KS-equation has been calculated
to be $\delta=0.230$ for the rescaled parameterless
version of the KS-equation \cite{Manneville85},
$\partial_t \phi = - \partial_x^2 \phi - \partial_x^4
\phi - \phi \partial_x \phi$.  Restoring the original
space, time, and magnitude scalings gives the
following~$c_1$ and~$c_3$ dependence of the dimension
density for KS~phase turbulence:
\begin{equation}
  \delta = 0.230
    \left(
      2 (c_1 c_3  -  1)  \over
      c_1^2 (1 + c_3^2)
    \right)^{1/2}
  . \label{ks-delta}
\end{equation}
In Fig.~\protect\ref{fig:delta-and-rms}(a), we compare
Eq.~(\protect\ref{ks-delta}) with our empirically determined values
of~$\delta$ for Eq.~(\protect\ref{1d-cgl}) from
Fig.~\protect\ref{fig:correlation-lengths}(b). The agreement is
good up to about $c_3 = 0.5$ ($\epsilon = .75$) and
then there is an increasing deviation of the actual
solutions from Eq.~(\protect\ref{ks-delta}).  This deviation with
increasing~$c_3$ may arise because the adiabatic
approximation Eq.~(\protect\ref{adiabatic-approximation}) breaks
down or because higher-order terms in the KS-equation
are renormalizing the dimension density.
Fig.~\protect\ref{fig:delta-and-rms}(b) gives some further
insight by comparing the mean-square fluctuation of
$\rho$ from Eq.~(\protect\ref{1d-cgl}) with the mean-square
fluctuation of~$\rho$ as given by
Eq.~(\protect\ref{adiabatic-approximation}).  We observe a
previously unreported power-law scaling of these
amplitude fluctuations with exponent~$\alpha=4$ from
the Newell line to near the $L_1$~line. Sufficiently
close to the Newell line, an exponent of~4 is predicted
by rescaling the solutions of Eq.~(\protect\ref{ks-eq}). The
adiabatic approximation is satisfied over a larger
range in~$c_3$ than the agreement between dimension
densities.

In conclusion, we have used a powerful parallel
computer to characterize more carefully the change from
defect- to phase-turbulence near the $L_1$~line in the
periodic one-dimensional Ginzburg-Landau equation in
the limit of large system size. Instead of a sharp
continuous transition with power-law scaling of order
parameters \cite{Shraiman92}, we found a smooth
crossover with new and anomalous structure near
the~$L_1$~line, e.g., the variation of~$\xi_\phi$ in
Fig.~\protect\ref{fig:correlation-lengths}(a) and the change in
slopes of dimension density and amplitude fluctuations
in Fig.~\protect\ref{fig:delta-and-rms}. We confirmed recent
predictions \cite{Shraiman92,Grinstein94} that the
phase correlation function decayed exponentially well
to the left of the $L_1$~line, with the related
correlation length being finite and large.  We also
found {\em two} length scales associated with the
field~$u$, a long scale associated with phase
fluctuations and a short scale~$\xi_\rho$ associated
with amplitude fluctuations.  Surprisingly, the
length~$\xi_\rho$ equals, up to a constant factor, the
dimension correlation length~$\xi_\delta$ associated
with the dimension density. This suggests that spatial
correlations of certain observables may suffice to
estimate big fractal dimensions of some large
homogeneous chaotic systems.

Numerous interesting questions remain for future study.
Our calculations leave open the theoretical question of
whether phase turbulence ($n_D = 0$) exists in the
thermodynamic limit. More generally, it is still not
known whether a chaos-to-chaos nonequilibrium
transition can occur in an infinite one-dimensional
system. Some of our results might be tested by
experiment \cite{cgl-expts}, e.g., the variation of
phase correlation length
(Fig.~\protect\ref{fig:correlation-lengths}(a)) and the scaling
of amplitude fluctuations near the Newell line
(Fig.~\protect\ref{fig:delta-and-rms}(b)).  It would be
interesting to extend our calculations to other parts
of the $c_1$-$c_3$ parameter plane, e.g., to understand
the hysteretic bichaotic regime $c_1 < 1.8$, for which
defect-turbulent and phase-turbulent states evidently
coexist in parameter space \cite{Shraiman92,ChateRefs}.
If phase-turbulence is just a small~$n_D$ limit of
defect-turbulence, it is more difficult to understand
the coexistence of different states of identical
symmetry. Finally, it would be interesting to repeat
similar calculations in two- and three-space
dimensions, for which the point-like space-time defects
in one-space dimension are replaced by long-lived
topological defects such as vortices and lines
\cite{Grinstein94}.

We thank T.~Bohr, L.~Bunimovich, H.~Chat\'e,
G.~Grinstein, C.~Jayaprakash, C.~O'Hern, J.~Sethna, and
B.~Shraiman for useful discussions and P.~Hohenberg and
J.~Socolar for numerous helpful comments on a
preliminary version of this paper. This work was
supported by grants from the National Science
Foundation and from the Department of Energy, and by
allotments of supercomputer time from the North
Carolina Supercomputing Center and from the National
Center for Supercomputing Applications.  The first
author would like to thank ONR for additional support.


\begin{references}
\bibitem{defect-phase-chaos-defns}
Defect-turbulence is a spatiotemporal chaotic state
defined by a finite density~$n_D$ of defects (measured
in defects per unit time per unit length), where a
defect is a space-time point~$(x,t)$ at which the field
amplitude~$\rho(x,t) = |u|$ vanishes and where the
phase $\phi(x,t)$ can slip by a multiple of~$2\pi$.
Phase turbulence is a chaotic state defined by the
absence of defects so that $n_D = 0$.
\bibitem{tau-defn}
The correlation time~$\tau$ of phase fluctuations was
obtained by identifying graphically the region of
asymptotic exponential decay $C(t) \propto
\exp(-t/\tau)$ of the correlation function~$C(t) =
\langle e^{i [\phi(x,t+t') - \phi(x,t')]} \rangle$
where $e^{i \phi} = u/|u|$ is defined numerically in
terms of a solution~$u$ of Eq.~(\protect\ref{1d-cgl}).  The angle
brackets denotes a space-time average over the region
$(t',x) \in [0,T]\times [0,L]$. Similarly, the phase
correlation length~$\xi_\phi$ was obtained by
identifying asymptotic exponentially decaying regions
of the two-point function $C(x) = \langle e^{i
[\phi(x+x',t) - \phi(x',t)]} \rangle$. Some correlation
function plots are given in Refs~\cite{Shraiman92}
and~\cite{Egolf94nature}.
\bibitem{xi-phi-scaling}
For the KS-equation in parameterless form, $\phi_t = -
\partial^2_x\phi - \partial^4_x\phi - (\partial_x
\phi)^2$ on a periodic interval of length~$L=512$, we
have found numerically that the correlation function of
$e^{i \phi(x,t)}$ \cite{tau-defn} decays exponentially
with correlation length~$\xi_\phi^{\text{KS}} = 9.6$.
The desired scaling relation then follows by using
Eqs~(4) and~(8) of Ref.~\cite{Shraiman92} to establish
a relation between the phase diffusion coefficients~$D$
of the KS- and CGL-equations. One finds that $\xi_\phi
= \xi_\phi^{\text{KS}} f(c_1,c_3) \epsilon^{-5/2}$ with
$f(c_1,c_3) = c_1 (c_1+c_3)^2 \sqrt{(1 + c_3^2)/2}$.
This expression for~$\xi_\phi$ is plotted in
Fig.~\protect\ref{fig:correlation-lengths}(a) for~$c_1=3.5$.
\end{references}

\newpage
\begin{figure}   
\caption{
{\bf (a)} Log-log plot of the space-time defect
density~$n_D$ versus the distance $c_3 - \bar{c_3}'$ to
the fitted point~$\bar{c_3}'$ where the density goes to
zero (onset of phase turbulence) for system size
$L=4096$, integration time~$T=10^5$, and an average
over~64 randomly specified initial conditions.  The
arrow labeled ``$L_1$'' indicates the position of the
$L_1$-line for parameter value~$c_1=3.5$
\protect\cite{Shraiman92}. The smallest~$n_D$ value
corresponds to a count of 200~defects. The two solid
lines were drawn to indicate the previous and present
best estimates of the exponent~$\alpha$ of a power-law
scaling. The crosses are the data from
Ref.~\protect\cite{Shraiman92}. In {\bf (b)}, we find
that Eq.~\protect\ref{essential-singularity} with
exponent~$\alpha=1$ gives a better fit of the same
data, with an onset of phase turbulence at $\bar{c_3}''
= 0.70$. The straight line is a plot of
Eq.~\protect\ref{essential-singularity} over the range
of its fit.  }
\label{fig:defect-density}
\end{figure}

\begin{figure}   
\caption{
{\bf (a)} Plot of the reciprocal square root of the
correlation time~$\tau$ (as estimated from the
asymptotic exponential decay of time correlation
functions of the phase~$e^{i\phi}=u/|u|$) versus the
parameter~$c_3$. The arrow indicates where the
$L_1$-line occurs for~$c_1=3.5$. We used a system size
of~$L=4096$, an integration time of~$T=10^5$, and an
average over 64~random initial conditions. The crosses
denote the similar data from
Ref.~\protect\cite{Shraiman92}.  {\bf (b)} For the same
numerical parameters, a plot of the product of
correlation time with defect density, $\tau n_D$,
showing that the scaling~$\tau \propto 1/n_D$ breaks
down near the $L_1$-line.  }
\label{fig:correlation-times}
\end{figure}

\begin{figure}   
\caption{
{\bf (a)} Plot of the phase correlation
length~$\xi_\phi$ for solutions of
Eq.~\protect\ref{1d-cgl} for system sizes of up
to~$L=10^6$, integration times of up to~$T=2\times
10^5$, and averages 64 randomly chosen initial
conditions. The crosses denote the similar data from
Ref.~\protect\cite{Shraiman92}. The solid curve is the
analytical expression obtained by scaling the finite
correlation length of the parameterless KS-equation
\protect\cite{xi-phi-scaling}. {\bf (b)} Plot of the
Lyapunov dimension density~$\delta$
\protect\cite{Egolf94nature} and the
reciprocals~$1/\xi_\rho$ and~$1/\xi_\phi$ of the
amplitude and phase correlation lengths. The reciprocal
length~$1/\xi_\rho$ (open circles) has been scaled by a
constant factor of~$0.7$ to emphasize the close
agreement with~$\delta$.  The positions of the Newell-
and $L_1$-lines for~$c_1=3.5$ are denoted by the arrows
labeled ``N'' and ``$L_1$'' respectively.  }
\label{fig:correlation-lengths}
\end{figure}

\begin{figure}   
\caption{
{\bf (a)} Comparison of the dimension density~$\delta$
for solutions of Eq.~(\protect\ref{1d-cgl}) with the rescaled
dimension density of the KS-equation, Eq.~(\protect\ref{ks-delta}),
for~$c_1=3.5$.  {\bf (b)} Comparison of the
mean-square fluctuations of the amplitude~$\rho$ as
calculated from the 1d CGL equation and as calculated
from the adiabatic approximation,
Eq.~(\protect\ref{adiabatic-approximation}), with~$\phi$ determined
from Eq.~(\protect\ref{1d-cgl}.  In both {\bf (a)} and~{\bf (b)}),
the arrows labeled ``N'' and ``$L_1$'' denote the
positions of the Newell- and $L_1$-lines respectively
for~$c_1=3.5$.  }
\label{fig:delta-and-rms}
\end{figure}

\end{document}